\documentclass[secnumarabic,footinbib,tightenlines,nobibnotes, aps, prb, twocolumn, superscriptaddress]{revtex4-1}
\usepackage{amsmath,amssymb}
\usepackage{graphicx}
\usepackage{dcolumn}
\usepackage{bm}
\usepackage{braket}
\usepackage{verbatim}
\usepackage{float}
\usepackage{bbold}
\usepackage{color}
\usepackage{xcolor}
\usepackage{relsize}

\usepackage{slashed}
\usepackage{amsthm}
\usepackage[T1]{fontenc}
\usepackage[colorlinks=true,urlcolor=blue,urlbordercolor={0 1 1}]{hyperref}

\begin{document}
\title{Paired exciton condensate and topological charge-$4e$ composite fermion pairing in half-filled quantum Hall bilayers}
\author{Ya-Hui Zhang}
\email{yahuizh@mit.edu}
\affiliation{Department of Physics, Massachusetts Institute of Technology, Cambridge, MA 02139, USA}
\author{Itamar Kimchi}
\affiliation{Department of Physics, Massachusetts Institute of Technology, Cambridge, MA 02139, USA}
\affiliation{Department of Physics, University of Colorado at Boulder, Boulder, CO 80309, USA}

\date{\today}

\begin{abstract}
Half-filled Landau levels admit the theoretically powerful fermion-vortex duality but longstanding puzzles remain in their experimental realization as $\nu_T=1$ quantum Hall bilayers, further complicated by Zheng et al's recent numerical discovery of an unknown phase at intermediate layer spacing.  Here we propose that half-filled quantum Hall bilayers ($\nu_T=1$) at intermediate values of the interlayer distance $d/\ell_B$ enter a phase with \textit{paired exciton condensation}.
This phase shows signatures analogous to the condensate of interlayer excitons (electrons bound to opposite-layer holes) well-known for small $d$ but importantly condenses only exciton pairs. 
To study it theoretically we derive an effective Hamiltonian for bosonic excitons $b_k$ and show that the single-boson condensate suddenly vanishes for $d$ above a critical  $d_{c1} \approx 0.95 l_B$. 
The nonzero condensation fraction $n_0=\langle b(0) \rangle ^2$ at $d_{c1}$ suggests that the phase stiffness remains nonzero for a range of $d>d_{c1}$ via an intermediate phase of paired-exciton condensation,  exhibiting $\langle bb \rangle \neq 0$ while $\langle b \rangle =0$.
Motivated by these results we derive a $K$-matrix  description of the paired exciton condensate's topological properties from composite boson theory. The elementary charged excitation is a half meron with $\frac{1}{4}$ charge and fractional self-statistics $\theta_s=\frac{\pi}{16}$. Finally we argue for an equivalent description via the $d=\infty$ limit through topological charge-$4e$ pairing of composite fermions. 
We suggest graphene double layers should access this phase and propose various experimental signatures, including an Ising transition $T_{Ising}$ below the Berezinskii-Kosterlitz-Thouless transition $T_{BKT}$ at $d \sim d_{c1}$.
\end{abstract}

\maketitle

\section{Introduction}

The half filled landau level, long a playground for exotic quantum Hall physics\cite{prange1987quantum}, has recently enjoyed a resurgence of theoretical interest. Perhaps the most remarkable theoretical connection has been to the newly discovered  fermion-vortex field theoretic duality\cite{son2015composite,metlitski2016particle,wang2015dual,wang2016half}. 
This duality has provided an alternate viewpoint on the composite-Fermi-liquid phase seen experimentally in the spin polarized $\nu=1/2$ lowest landau level, presenting the composite fermion that forms this state as the vortex-dual of the electron\cite{son2015composite,wang2016half}.

Though measurable manifestations of the new theories have been difficult to come by for a single half-filled landau level, the outlook is much more promising for a pair of coupled half-filled landau levels, as realized in a bilayer quantum Hall  system. Quantum Hall bilayers\cite{eisenstein2014exciton} show interesting behavior as the total filling factor $\nu_\text{T} = \nu_1 + \nu_2$ is tuned near the value $\nu_\text{T} = 1$ which corresponds to a $\nu_1=\nu_2=1/2$ half filled Landau level in each layer.\footnote{The phases discussed below are all stable to small imbalances that deviate each layer from half filling.} When the distance $d$ between the layers is sufficiently small (compared to the magnetic length $\ell_B$), $\nu_\text{T} =1$ bilayers exhibit a sharp peak in the interlayer tunneling conductance and a vanishing counterflow Hall resistance, providing strong evidence for an \textit{exciton condensate} phase\cite{suen1992observation,eisenstein1992new}.
\footnote{The exciton condensate phase spontaneously breaks a U(1) symmetry in the limit of vanishing interlayer tunneling; for theoretical concreteness we take this limit throughout this manuscript.} 
This phase is simply understood as the Bose-Einstein Condensation (BEC) of the bosonic exciton bound state of an electron from one layer and a hole from the other. Interestingly, the conventional signatures of this phase disappear as the layers are taken farther apart near $d/\ell_B\approx 2.0$. Despite decades of work, this observed behavior with increasing $d$ has not yet found any commonly accepted interpretation.

Observations of the exciton condensation phase in quantum well experiments begin at $d\gtrsim 1.5 \ell_B$. In $d\rightarrow 0$ limit, exciton condensation is theoretically well established through simple Hartree Fock calculation\cite{sondhi1993skyrmions,moon1995spontaneous,yang1996spontaneous}. In agreement with the theory, the exciton condensation phase is indeed observed in graphene system in the region $d \lesssim 0.8 \ell_B$\cite{li2017excitonic}. Given these experimental results, one may expect that the same exciton condensation phase extends from small $d$ to $d\approx 2.0 \ell_B$.  The scenario of an extended exciton condensate phase is further supported by theoretical results that the same phase is also the likely fate of the system in the $d\rightarrow \infty$ limit\cite{sodemann2017composite} if disorder is theoretically taken to be much smaller than the small gap.\footnote{In the limit of infinite distance, the low energy theory contains two decoupled Composite Fermion Liquid (CFL). Using the Dirac theory of the CFL, Ref.~\onlinecite{sodemann2017composite} shows that small inter-layer repulsion leads to an instability of an inter-layer pairing of composite fermions. The resulting phase, assuming zero-angular-momentum pairing of Dirac CFs, is shown to be topologically equivalent to an exciton condensation phase.} Theoretically the exciton condensation phase at  small $d$ can cross over (in a manner similar to the BEC-BCS crossover) to the inter-layer pairing phase of composite fermions at large $d$ at least for some Hamiltonians, and the disappearance of the exciton condensation phase in the quantum well experiments after $d\gtrsim 2.0 \ell_b$ can be explained by disorder effects, seemingly consistent with a single exciton condensate phase spanning the entire $d$-dependent phase diagram.

However, recently numerical simulations of realistic Hamiltonians uncovered evidence against this simple scenario.  A DMRG calculation of lowest-landau-level projected unscreened Coulomb interactions found a level crossing at a critical $d=d_{c1}$ with $d_{c1}\approx 1.1 \ell_B$ and an ED calculation confirmed the level crossing and further showed that the system enters an intermediate phase above  $d\gtrsim 1.1 \ell_B$\cite{zhu2017numerical}. In the intermediate phase, the phase stiffness for the exciton condensation was seen to remain finite. However, the single exciton gap (the gap for moving one electron from one layer to the other layer) was found to extrapolate to a finite value by finite-size scaling, while the two exciton gap was extrapolated to zero. There are several theoretical proposals for intermediate phases in quantum Hall bilayers\cite{alicea2009interlayer,you2017interlayer,lian2018wave}, but as far as we are aware, none of these theories are consistent with the numerical result of Ref.~\onlinecite{zhu2017numerical}.

Motivated by the experimental puzzles together with the numerical observation in Ref.~\onlinecite{zhu2017numerical}, here we propose a \textit{paired}-exciton condensation phase at intermediate $d$ in $\nu_T=1$ quantum Hall bilayers. The proposed paired-exciton condensate shows experimental signatures analogous to the single exciton condensate and can explain the existing numerical results by Ref.~\onlinecite{zhu2017numerical}. We describe how to connect the paired exciton phase to both $d\rightarrow 0$ and $d\rightarrow \infty$ limits. Studying this phase we find a half-meron excitation with fractional charge $Q=\frac{e}{4}$ and fractional self statistics $\theta_s=\frac{\pi}{16}$, which may be further tested by future numerical simulations. In addition, we argue that close to $d \lesssim d_{c1}$, there is a finite temperature Ising transition separating the single exciton condensation (EC) phase and the paired exciton condensation (PEC) phase at $T_c$ below the BKT transition at $T_{BKT}$. Meanwhile we expect that $T_{BKT}$ shows particular behavior near $d=d_{c1}$, with a steep decrease in the value of $T_{BKT}$ but a slightly positive second derivative $\partial^2 T_{BKT}/ \partial d^2 \gtrsim 0$. The various predictions can be tested by  future experiments in graphene double layers at larger interlayer distance. 

At larger $d$, we expect a single exciton condensation phase from the pairing instability of the Composite Fermion Liquids (CFL) of the two layers\cite{sodemann2017composite}. Therefore we expect the paired exciton condensation phase is sandwiched by two single exciton condensation phases with two critical point $d_{c1}$ and $d_{c2}$.  Theoretically both transitions can be continuous. $d_{c1}$ is associated with superfluid to paired superfluid transition for composite bosons. We argue that the transition at $d_{c2}$ can be understood as the transition from charge $2e$ pairing to charge $4e$ pairing for composite fermions.   As we can not  determine the value of $d_{c2}$, it is also plausible that the observations of exciton condensation in quantum well experiments arise from a paired exciton condensation phase rather than a single exciton phase, and single exciton condensation would then occur only at smaller layer separations.  Settling this question would require  experimental data in the intermediate region; casual readers may skip ahead to the end of this manuscript, where we give a summary of experimental signatures of the paired exciton condensate phase.

\section{Effective Theory for Excitons}

We consider a bilayer quantum Hall system with number of electrons $N_1=N_2=N=\frac{N_\Phi}{2}$ where $N_{\Phi}$ is the number of fluxes. We assume the number of particles for each layer is separately conserved. Projected to the Lowest Landau Level (and ignoring any interlayer tunneling in the Hamiltonian), the Hamiltonian of the problem arises purely from the Coulomb interaction: 
\begin{equation}
  H=\frac{1}{2}\sum_{a,b=1,2}\frac{1}{(2\pi)^2}\int d^2 q \tilde V_{ab}(\mathbf q) \tilde \rho_a(\mathbf q) \tilde \rho_b(-\mathbf q)
\end{equation}
where $a,b$ are layer indexes.
\begin{equation}
  \tilde V_{aa}=\frac{1}{q}e^{-\frac{q^2}{2}}
\end{equation}
and
\begin{equation}
 \tilde  V_{12}=\tilde V_{21}=\frac{1}{q}e^{-\frac{q^2}{2}}e^{-qd}
\end{equation}

Next we define the Hilbert Space. For simplicity, we use electron operator $c_m$ for layer $1$ and hole operator $f_n$ for layer $2$. Here $m,n=1,2,...,N_\Phi$ is the index for the Landau orbitals.  Any state in the Hilbert space can be written as:
\begin{align}
  \Psi&=\sum g(m_1,...,m_N;n_1,...,n_N)c^\dagger_{m_1} f^\dagger_{n_1}...c^\dagger_{m_N} f^\dagger_{m_N} \ket{\tilde 0}\notag\\
      &=\sum g(m_1,...,m_N;n_1,...,n_N) \epsilon_{n_1,...,n_N} b^\dagger_{m_1n_1}...b^\dagger_{m_Nn_N}\ket{\tilde 0}
\label{eq:psi_b}
\end{align}
where the vacuum $\ket{\tilde 0}$ is defined as the Integer Quantum Hall Insulating state with all Landau levels filled for layer $2$. We write $b^\dagger_{mn}=c^\dagger_m f^\dagger_n$.  $g(m_1,...,m_N;n_1,...,n_N)$ is a symmetric function under interchange of $m_i,m_j$ or $n_i,n_j$. $\epsilon_{m_1,...,m_N}$ is the antisymmetric tensor to enforce the correct commutation relations for fermions. $b_{mn}$ is therefore a boson formed as a bound state of an electron of layer $1$ and a hole of layer $2$.

In first quantization language,  from $\Psi$ in the above equation we can construct a wavefunctions for electrons $z_i$ in layer $1$ and holes $\bar w_{N+i}$ in layer $2$ as $\Psi_b(z_1,\bar w_{N+1};...;z_N;\bar w_{2N})$. We can then obtain a wavefunction of electrons for both layers through the following projection:
\begin{align}
  &\Psi(z_1,...,z_N;w_1,...,w_N)=\notag\\
  &\ \int dw_{N+1}...dw_{2N} \Psi^*_{IQHE}(w_1,...,w_{2N})\Psi_b(z_1,\bar w_{N+1};...;z_N;\bar w_{2N})  
  \label{eq:projection}
\end{align}
where $z$ is coordinate for first layer and $w$ for second layer.  $\Psi_{IQHE}$ is the wavefunction for the fully filled state:
\begin{equation}
  \Psi_{IQHE}(w_1,...,w_{2N})=\prod_{j>i}(w_j-w_i)
\end{equation}

Therefore any state in the Hilbert space can be mapped from $\Psi_b$, which is constructed from bosonic operator $b_{m_i n_j}$. $\epsilon_{n_1,...,n_N}$ tensor in Eq.~\ref{eq:psi_b}  or equivalently the projection in Eq.~\ref{eq:projection} provide the constraint that the fundamental  degree of freedom is fermionic.  Generally such a constraint is hard to enforce in analytical calculation. However, at the small distance limit, fermionic degree of freedom is gapped and the low energy theory should be just a theory of bosonic excitons. Therefore we can make an approximation to ignore the $\epsilon_{n_1,...,n_N}$ constraint in Eq.~\ref{eq:psi_b} and write down effective theories for bosonic operators $b_{mn}$. Such an approximation should be valid as long as the gap to break this bosonic bound state is finite, which is true for the exciton condensation phase and the paired exciton condensation phase we propose.

In the following we are going to derive an effective Hamiltonian for these bosonic excitons with operator $b_{mn}$. First, we can replace the Landau index $m,n$ with a well-defined center-of-mass momentum $\mathbf{k}$. 
Each state $b_{\mathbf k}^\dagger$ creates the following wave-function for one single exciton:
\begin{align}
  \braket{z,\bar{w}|\mathbf k}&=\tau_{\mathbf k}(z,\bar{w})\notag\\
  &=\frac{1}{2\pi}e^{-\frac{1}{4}|z|^2-\frac{1}{4}|w|^2+\frac{1}{2}z\bar{w}}e^{\frac{i}{2}(\bar{k}z+k\bar{w})}e^{-\frac{1}{4}|k|^2}\notag\\
  &=\frac{1}{2\pi}e^{-\frac{1}{4}|\mathbf{r}-\wedge{\mathbf{k}}|^2}e^{i\mathbf{k}\cdot \mathbf{R}}e^{-\frac{i}{2}\mathbf{r}\wedge \mathbf{R}}
\end{align}
where $\wedge{\mathbf k}=\hat{z}\times \mathbf{k}$, $\mathbf{R}=\frac{\mathbf{z}+\mathbf{w}}{2}$ and $\mathbf{r}=\mathbf{z}-\mathbf{w}$.
It can be proven\cite{pasquier516dipole,read1998lowest} that (see Appendix.~\ref{appendix:exciton_eigen_states}):
\begin{equation}
  \sum_{mn}\ket{mn}\bra{mn}=\sum_{\mathbf k}\ket{\mathbf k}\bra{\mathbf k}
  \label{eq:complete_condition}
\end{equation}

One can see that each $b^\dagger_{\mathbf k}$ creates a bosonic exciton with center of mass momentum $\mathbf k$. Besides, the exciton has an internal dipole moment $\mathbf{P}=\wedge{\mathbf k}$. This dipole moment gives the kinetic term for excitons because of the inter-layer attractive Coulomb interaction between electron and hole.
\begin{equation}
  H_K=\sum_{\mathbf k}\xi_{\mathbf k}b^\dagger_{\mathbf k} b_{\mathbf k}
  \label{eq:kinetic}
\end{equation}
with
\begin{align}
  \xi_k &=-\frac{1}{2\pi} \int d^2 r \frac{1}{\sqrt{r^2+d^2}}e^{-|\mathbf{r}-\wedge \mathbf k|^2}\notag\\
  &=-\frac{1}{2\pi} \int dr d\theta  \frac{r}{\sqrt{r^2+d^2}}e^{-(r^2+k^2-2kr\cos\theta)}\notag\\
  &=- \int dr   BesselI[0,2kr] \frac{r}{\sqrt{r^2+d^2}}e^{-(r^2+k^2)}\notag\\
\end{align}
where $-\frac{1}{2\pi}\frac{1}{\sqrt{r^2+d^2}}$ is the attractive Coulomb interaction between a electron at layer $1$ and a hole with layer $2$ with distance $r$ in $xy$ plane.

We simulated the above equation numerically. At the $k<1$ limit, we find $\xi(k)=\frac{k^2}{2m}+\xi(0)$ with $m\approx1.135+6.187 d^2$.

Density-density interaction terms are introduced by
\begin{equation}
  H_V=\frac{1}{2}\sum_{a,b=1,2}\frac{1}{(2\pi)^2}\int d^2 \mathbf{q}  V_{ab}(\mathbf q)  \rho_a(\mathbf q)  \rho_b(-\mathbf q)
  \label{eq:ohm}
\end{equation}
with 
\begin{equation}
  \rho_1(\mathbf q)=\sum_{\mathbf k} b^\dagger_{\mathbf{k-q}}b_{\mathbf k} e^{\frac{i}{2}\mathbf{k}\wedge \mathbf{q}}
\end{equation}
and
\begin{equation}
  \rho_2(\mathbf q)=\sum_{\mathbf k} b^\dagger_{\mathbf{k-q}}b_{\mathbf k} e^{-\frac{i}{2}\mathbf{k}\wedge \mathbf{q}}
\end{equation}
where $\mathbf{k}\wedge \mathbf{q}$ denotes the cross product. We can easily check that the following Girvin-Macdonald-Platzman (GMP) algebras are satisfied: 

\begin{align}
&[\rho_1(\mathbf{q_1}),\rho_1(\mathbf{q_2})]=2 i\sin \frac{\mathbf{q_1}\times \mathbf{q_2}}{2} \rho_1(\mathbf{q_1+q_2})\notag\\
&[\rho_2(\mathbf{q_1}),\rho_2(\mathbf{q_2})]=-2 i\sin \frac{\mathbf{q_1}\times \mathbf{q_2}}{2} \rho_2(\mathbf{q_1+q_2})
\end{align}

Following some simple algebra, the four-particles interaction for $b$ is
\begin{align}
H_V&=\int \frac{d\mathbf k_1^2}{(2\pi)^2}\frac{d \mathbf k_2^2}{(2\pi)^2}\frac{d \mathbf q^2}{(2\pi)^2}b^\dagger_{k_1-q}b_{k_1}b^\dagger_{k_2+q}b_{k_2}\frac{1}{q}e^{-\frac{q^2}{2}}\notag\\
&\left(\cos\left(\frac{(\mathbf k_1-\mathbf k_2)\wedge \mathbf q}{2}\right)-e^{-qd}\cos\left(\frac{(\mathbf k_1+\mathbf k_2)\wedge \mathbf q}{2}\right)\right)
\label{eq:four-boson}
\end{align}

$H=H_{K}+H_V$ with $H_K$ in Eq.~\ref{eq:kinetic} and $H_V$ in Eq.~\ref{eq:four-boson} form the effective Hamiltonian for the  exciton bosons.

\section{Bogoliubov Mean Filed Theory For Exciton Condensation Phase}
At the $d<<1$ limit, one has a smaller mass $m$ and the interaction is almost cancelled. Therefore we expect that bosons condense to the $k=0$ state which results in $\langle b \rangle \neq 0$.

Let us now give a mean field theory calculation of this exciton condensation phase, following the standard Bogoliubov theory for the weakly interacting bosons.  We assume $\langle b \rangle=\Phi=\sqrt{n_0}$. Because of the interaction, $\frac{n_0}{n}$ is finite but generically not equal to $1$.  $n_0$ can be decided from the following self consistent equation(see Appendix.~\ref{appendix:self-consistent-equation}):
\begin{equation}
  n-n_0=\frac{1}{2V}\sum_{\mathbf q \neq 0}\frac{\xi_{\mathbf q}+\Delta_{\mathbf q}-E_{\mathbf q}}{E_{\mathbf q}}
  \label{eq:self_mean}
\end{equation}
where $\Delta_{\mathbf q}=2 n_0 V_{+}(\mathbf q)$ with $V_{+}(\mathbf q)=\frac{e^{-\frac{|q|^2}{2}}}{|q|}(1-e^{-|q|d})$.
We can solve Eq.~\ref{eq:self_mean} self consistently and get the condensation percentage $\frac{n_0}{n}$.

\begin{figure}[ht]
\centering
\includegraphics[width=250pt]{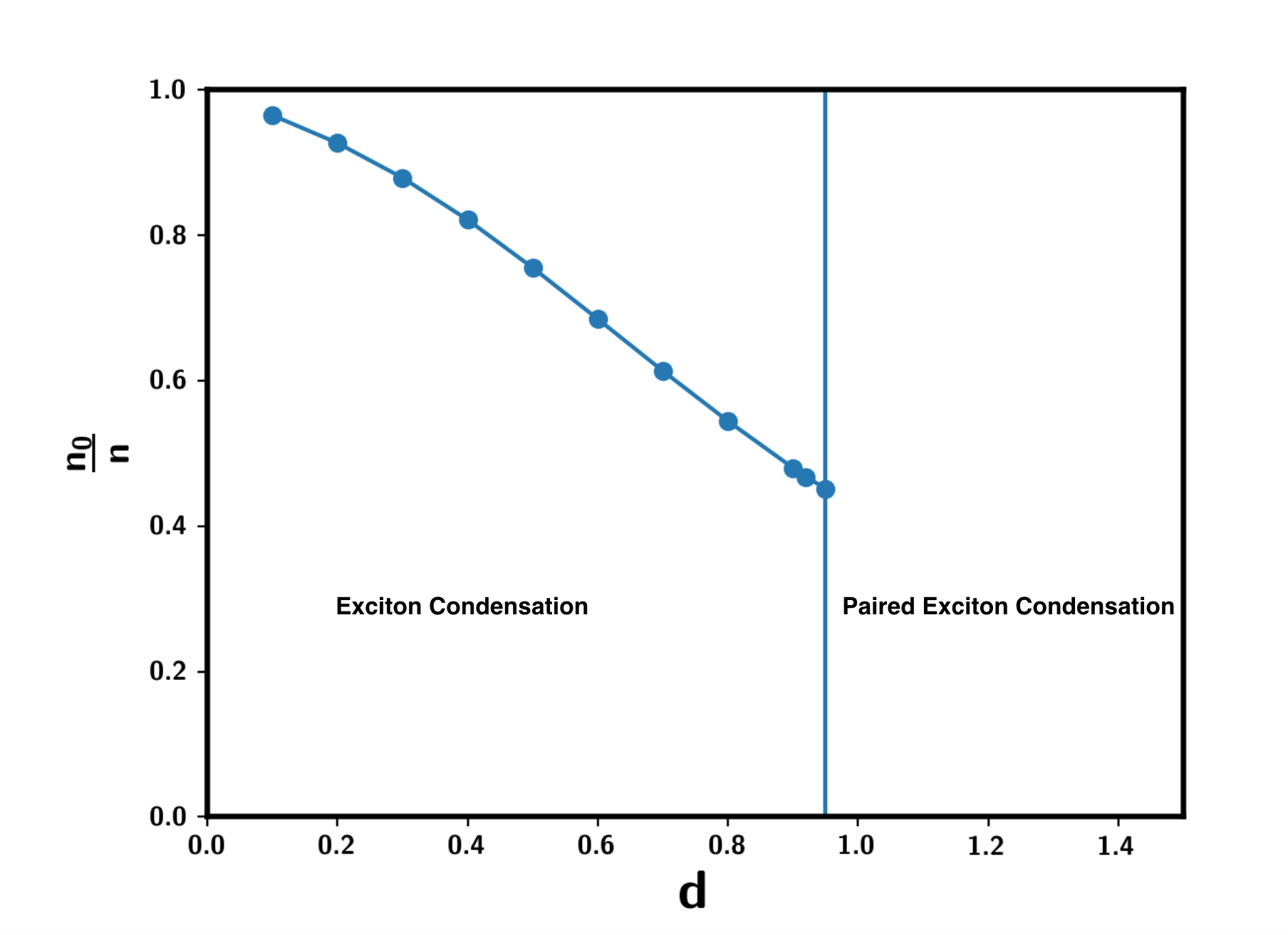}
\caption{Self-consistent solutions for the single-exciton condensate fraction $\frac{n_0}{n}$ as a function of interlayer distance $d$ (in units of $\ell_B$). 
No solution of the  single-boson condensate can be found for  $d > d_{c1} \approx 0.95 \ell_B$. We expect the phase entered at $d_{c1}$ is a paired superfluid of the bosonic excitons. Note that this $d_{c1}\approx 0.95 \ell_B$ is close to the numerical value $d_{c1}\approx 1.1 \ell_B$ of Ref.~\onlinecite{zhu2017numerical}. The phase transition can be a continuous Ising transition, though it is difficult to describe it with a simple mean field theory. }
\label{fig:nd}
\end{figure}

As shown in Fig.~\ref{fig:nd}, for $d>d_{c1}$, with $d_{c1} \approx 0.95$, there is no longer any solution with single-boson condensation.  One possibility is that the gap for fermionic excitation is closed and the above effective theory is not valid anymore. However, numerical result in Ref.~\cite{zhu2017numerical} suggests that phase stiffness for exciton condensation is finite as long as $d<1.8 \ell_B$. Therefore we expect that low energy is still dominated by bosonic excitons.  In this paper we assume this is true and then give a discussion about the possibility after $d>d_{c1}$.

Impossibility to find a solution for Eq.~\ref{eq:self_mean} means that chemical potential can not be set to be zero anymore. A non zero chemical potential $\mu\neq 0$ needs to be added and opens a gap for the single boson excitation. At $d_{c1}$, $\frac{n_0}{n}\approx 0.45$ means that there is still finite superfluid phase stiffness $\rho_s$. At $d>d_{c1}$, we expect a finite region where phase stiffness still remains finite while single boson is gapped. This naturally leads to the proposed paired-superfluid phase.

To describe a paired superfluid, one should add a non-zero chemical potential $\mu$. Then we need to decouple the original four boson interaction to  $A(\mathbf k) b^\dagger_{\mathbf k} b_{\mathbf k}$ and $B(\mathbf k)b_{\mathbf k}b_{-\mathbf k}$. The self consistent euqations for $\mu$, $A(\mathbf k)$ and $B(\mathbf k)$ are described in the Appendix.~\ref{appendix:self-consistent-equation}. The solution is hard to obtain, though we expect the existence of a solution when $d$ is only slightly larger than $d_{c1}\approx 0.95 \ell_B$.  In the following we will assume such a paired exciton condensation phase and discuss its topological properties based on composite bosons from flux attachment.

\section{Chern Simons Theory for Paired Exciton Condensation Phase}
We derive the Chern Simons theory for the paired exciton condensation phase here. By attaching the same flux to each original fermion for both layers, we can have an effective theory in terms of composite bosons\cite{zhang1992chern}:
\begin{align}
  S&=\sum_\sigma \int d^3x b^\dagger_\sigma\frac{(\partial_\mu-\tilde a_\mu-A^a_\mu)^2}{2m}b_\sigma\notag\\
  &+\frac{1}{4\pi}\int d^3x  \tilde a d \tilde a +\int d^3x d^3y \rho(x)V(x-y)\rho(y)
\end{align}
where pseudospin $\sigma=1,2$ is the layer index. The total flux $\frac{d\tilde a}{2\pi}=-(\rho^b_1+\rho^b_2)$ can fully cancel the background magnetic field.

When $d$ is small, composite boson condense: $\langle b_1 \rangle \neq 0$ and $\langle b_2 \rangle \neq 0$.  The following Effective theory can be generated using standard boson-vortex duality in $2+1$d\cite{dasgupta1981phase,lee1990boson}:
\begin{align}
L=\frac{1}{4\pi}\tilde a d\tilde a +\frac{(\tilde a+A_1)da_1}{2\pi}+\frac{(\tilde a+A_2)da_2}{2\pi}
\end{align}

Integrating $\tilde a $, we have
\begin{equation}
  L=-\frac{(a_1+a_2)d(a_1+a_2)}{4\pi}+\frac{A_1 d a_1}{2\pi}+\frac{A_2 da_2}{2\pi}
\end{equation}
which gives the well-known $K$ matrix theory for  exciton condensation phase\cite{wen1992neutral}.

At intermediate distance $d>d_c$, to describe the paired exciton condensation phase, we consider the case that single composite boson is gapped while a pair of composite bosons condenses: $\langle b_1b_1 \rangle \neq 0$ and $\langle b_2 b_2 \rangle \neq 0$.  We will show the properties of this phase match  those of the paired exciton condensation phase.

Similar to the single composite boson condensation, the effective theory from boson-vortex duality of the two pairings is
\begin{align}
L=\frac{1}{4\pi}\tilde a d\tilde a +\frac{2(\tilde a+A_1)da_1}{2\pi}+\frac{2(\tilde a+A_2)da_2}{2\pi}
\end{align}

After integrating $\tilde a$, we get
\begin{equation}
  L=-\frac{4(a_1+a_2)d(a_1+a_2)}{4\pi}+\frac{2A_1 d a_1}{2\pi}+\frac{2A_2 da_2}{2\pi}
\end{equation}

This corresponds to $K$ Matrix $K =\left(\begin{array}{cc}4&4\\4&4\end{array}\right)$\cite{wen2004quantum}. In this convention, excitations are labeled as $(l_1,l_2)$ with charge $l_i$ under $a_i$, $i=1,2$. The elementary excitation is $(1,0)$ and $(0,1)$.

For simplicity, we rewrite the Lagrangian in terms of $a^\mu_c=\frac{a^\mu_1+a^\mu_2}{2}$ and $a_s=\frac{a^\mu_1-a^\mu_2}{2}$,  $A^\mu_c=\frac{A^\mu_1+A^\mu_2}{2}$ and $A_s=\frac{A^\mu_1-A^\mu_2}{2}$. The convention of $A^{c}$ and $A^s$ is chosen to make their charges $q_c=q_1+q_2$ and $q_s=q_1-q_2$ for charge $q_1$ under $A_1$ and charge $q_2$ under $A_2$. In this convention, we have:

\begin{equation}
 L= -\frac{16}{4\pi}a_cda_c+\frac{4}{2\pi}A_cda_c+\frac{4}{2\pi}A_sda_s
 \label{eq:composite_boson_theory}
\end{equation}

Elementary excitation is still $(l_1,l_2)=(1,0),(0,1)$, or $(l_c,l_s)=(1,1),(1,-1)$. The corresponding physical charges are
\begin{align}
Q_c&=4\frac{da_c}{2\pi}\notag\\
Q_s&=4\frac{da_s}{2\pi}
\end{align}

Let's consider pseudo-spin excitation first. Because there is no Chern-Simons term for $a_s$, the pseudo-spin excitation has gapless modes. the smallest charge under $a_s$ is $l_s=1$ and therefore smallest flux of $a_s$ is $2\pi$. As a result, smallest gapless spin excitation is $Q_s=Q_1-Q_2=4\frac{da_s}{2\pi}=4$.  The single exciton corresponds to $Q_s=2$ and is gapped in this phase. This result is consistent with a paired exciton condensation phase and can explain the numerical result in Ref.~\onlinecite{zhu2017numerical}.

$l_c=1$ will induce charge $Q_c=\frac{1}{4}$. This can be understood as the $2\pi$ vortex of $<b_1b_1>$ and therefore a $\pi$ vortex of $b_1$, or $b_2$, which is a half meron. Therefore, in this phase $(l_1,l_2)=(1,0),(0,1)$ actually correspond to half meron and half anti-meron spin texture.  
With the Chern-Simons theory, it is also easy to get the self statistics of this half-Meron particle to be $\theta_s=\frac{\pi}{16}$\cite{wen2004quantum}.

All of the above properties are consistent with the paired exciton condensation phase, for which $Q_s=2$ is gapped while there are $Q_s=4$ gapless spin wave excitations. $Q_c=\frac{1}{4}$ of half-meron is easily understood following Laughlin's argument. 

At $d_{c1}$, the transition between the exciton condensation phase and the paired exciton condensation phase can be described by the superfluid to paired suplerfluid transition of composite bosons.

\begin{figure}
\centering
\includegraphics[width=250pt]{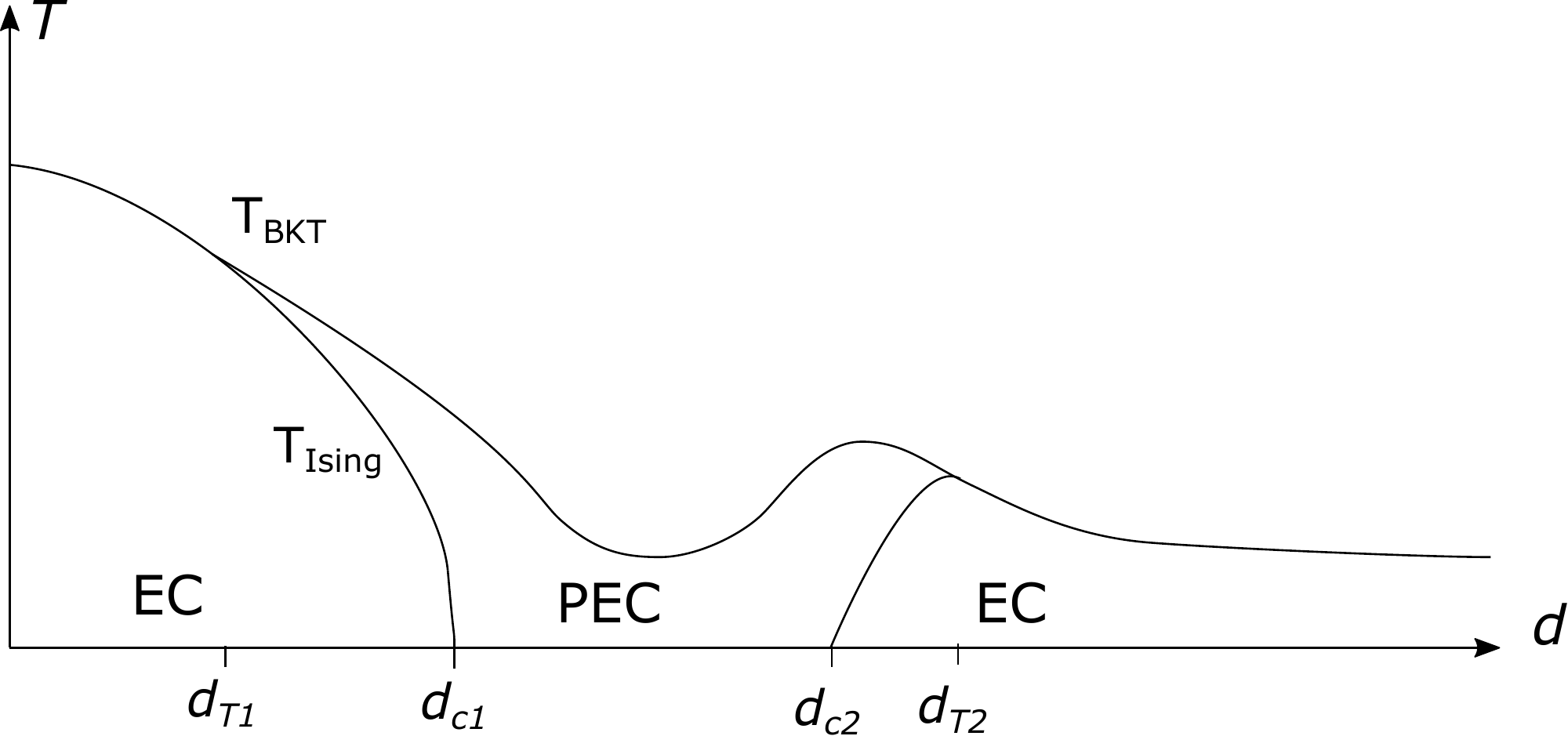}
\caption{Phase diagram for the clean quantum Hall bilayers with distance $d$. We expect a paired exciton condensation (PEC) phase sandwiched by two phases of single exciton condensation (EC). The two EC phases are topologically equivalent. A recent numerical result\cite{zhu2017numerical} suggested $d_{c1}\approx 1.1 \ell_B$ but $d_{c2}$ is fully unknown. Close to the quantum phase transitions at $d_{c1}$ and $d_{c2}$, we expect a thermal Ising transition from EC to PEC at $T_{Ising}$, below the BKT transition at $T_{BKT}$. The onset of these thermal Ising transitions is labeled $d_{T1}$ and $d_{T2}$. 
Note that the temperature axis is not drawn to scale, and in particular the low-$T$ transition of the large-$d$ EC phase may render it indistinguishable from compressible CFL phases upon adding disorder.}
\label{fig:phase_diagram}
\end{figure}

\section{Charge-$4e$ Pairing of dual Composite Fermions}
In the previous section we described the paired exciton condensation phase from the view of a composite boson theory; here we explore a description in terms of dual variables.  The bilayer system can be viewed as two layers of half-filled Landau levels. When interlayer distance $d$ is infinite, the two layers decouple and the system  is known to be described by two composite fermi liquids (CFLs). In terms of composite fermions, it has been well established that  the single exciton condensation phase corresponds to a  $p+ip$ pairing in the HLR picture (or $s$ wave pairing in the Dirac picture) of the composite fermions in triplet $S^z=0$ channel\cite{sodemann2017composite}. Then the following question arises: Is there a description of  the paired exciton condensation phase starting from two decoupled CFLs? In the following we will argue that the paired exciton condensation phase can be described by a topological charge $4e$ pairing of composite Fermions.

For simplicity, we first use the Dirac picture of CFLs following the discussion in Ref.~\onlinecite{sodemann2017composite}:
\begin{equation}
  L=\sum_{I=1,2}\bar \Psi_I(i\slashed \partial+\slashed a_I)\Psi_I+\frac{1}{4\pi} A_I d a_I +\frac{1}{8\pi}A_I d A_I
\end{equation}
where $I$ is the layer index, which can be viewed as a pseudospin index.

 To get an insulating state with gapless pseudospin excitation, we need to Higgs $a_c=\frac{a_1+a_2}{2}$ while keeping $a_s=\frac{a_1-a_2}{2}$ gapless. This requires inter-layer pairing of composite fermions. In a charge-$2e$ pairing of composite fermions, the $h/2e$ vortex of the pairing carries physical charge $1/2$ and corresponds to the meron excitation of the single exciton condensation phase\cite{sodemann2017composite}. To have a paired exciton condensation phase, we need a half meron carrying physical charge $\frac{1}{4}$. Therefore it is necessary to have a deconfined $h/4e$ vortex under $a_c$, implying a charge $4e$ pairing of composite fermions\footnote{Note that here we use ``charge $4e$ pairing'' as usual to denote pairing of four fermions, even though the Dirac composite fermions are electrically neutral.}.  To match the self statistics $\theta_s=\frac{\pi}{16}$ from the composite boson theory, the charge $4e$ superconductor needs to be topological. For the single exciton condensation, the meron also has a fractional self statistics $\theta_s=\frac{\pi}{4}$, which can be matched by the topological $2e$ pairing of composite fermions in the $\nu_{kitaev}=2$ class of  the Kitaev's 16 fold way classification\cite{kitaev2006anyons}.  Therefore we need a theory of a topological charge $4e$ superconductor to fully establish the connection between the paired exciton condensation phase and the charge $4e$ pairing of the composite fermions.

  Despite the lack of a mean field description for such a theory, Ref.~\onlinecite{wang2016braiding} classifies charge $4e$ superconductors based on a pure algebraic theory of topological orders. Interestingly the smallest possible self statistics of the $h/4e$ vortex is $\frac{\pi}{16}$, which matches the half-meron self statistics of the paired-exciton condensate from composite boson theory.  Aside from this fairly formal result, it is hard to figure out the form of the charge $4e$ pairing  corresponding to this topological class.  We leave it to future work to have a detailed description of this charge $4e$ pairing. Here we just quote this result and assume that the composite fermions are paired in this topological class. Then the resulting $h/4e$ vortex has the same topological property as the half-meron excitation in the $K$ matrix description of the paired exciton condensation phase. Encouraged by the above agreement, we conjecture that the same paired exciton condensation phase studied in the previous section can be equivalently described by a charge $4e$ pairing of the Dirac composite fermions. Checking this conjecture will require a more complete theory of the topological charge $4e$ superconductor of Dirac fermions.

  The above description is based on the Dirac theory for the CFLs. Next we show that the $K$ matrix theory in the previous section can be recovered by a topological charge $4e$ pairing of composite fermions in the Haperin-Lee-Read (HLR) theory\cite{halperin1993theory}.
  We start from two decoupled HLRs\footnote{To be more rigorous, we use a modified version of the HLR theory}:
  \begin{align}
    L&=L[c_1,a_1]+L[c_2,a_2]\notag\\
    &-\frac{2}{4\pi}\alpha_1 d \alpha_1+\frac{1}{2\pi}(A_1-a_1)d\alpha_1\notag\\
    &-\frac{2}{4\pi}\alpha_2 d \alpha_2+\frac{1}{2\pi}(A_2-a_2)d\alpha_2
    \label{eq:cfls}
  \end{align}

Then we consider a charge $4e$ pairing $\langle c_1 c_1 c_2 c_2 \rangle \neq 0$ in the topological class which gives the $h/4e$ vortex self-statistics $\theta_s=\frac{\pi}{16}$\cite{wang2016braiding}. According to Ref.~\onlinecite{wang2016braiding}, this topological order is Abelian. Therefore it should be described by a Chern-Simons theory with $K=16$. We use the following action for the topological charge $4e$ superconductor formed by the composite fermions: 
\begin{equation}
  L_{4e SC}=\frac{16}{4\pi}\beta_s d \beta_s+\frac{2}{2\pi}(a_1-a_2)d\beta_s+\frac{2}{2\pi}(a_1+a_2)d\beta_c
  \label{eq:charge_4e_sc}
\end{equation}
where $\beta_c$ is dual to the charge $4e$ pairing. $\beta_s$ is introduced to impose the topological property. We conjecture that this charge $4e$ superconductor has the same spin Hall effect as the $p+ip$ triplet $2e$ pairing and introduce the mutual Chern-Simons term $(a_1-a_2)d\beta_s$. Integrating $a_1$ and $a_2$, we can identify $\alpha_1=2(\beta_c+\beta_s)$, $\alpha_2=2(\beta_c-\beta_s)$.  Substituting these two relations into Eq.~\ref{eq:cfls} and Eq.~\ref{eq:charge_4e_sc}, we have the final theory for the state resulting from the charge $4e$ pairing:
\begin{equation}
  L=-\frac{16}{4\pi}\beta_c d \beta_c+\frac{4}{2\pi}A_c d \beta_c+\frac{4}{2\pi} A_s d \beta_s
\end{equation}
which is identical to Eq.~\ref{eq:composite_boson_theory} from the composite boson theory.

Although we do not have a mean field description of the charge $4e$ superconductor described by Eq.~\ref{eq:charge_4e_sc}, the agreement with the composite boson theory is a strong support for its existence. Therefore we believe that the paired exciton condensation phase from the paired superfluid of composite bosons can be equivalently 	constructed from the topological charge-$4e$ pairing of the composite fermions.														

  Next we discuss the transition between the exciton condensation and the paired exciton condensation in the composite fermion picture. As argued by Ref.~\onlinecite{sodemann2017composite}, at large distance the instability of two CFLs is an inter-layer charge $2e$ pairing which, in the simplest angular momentum channel, gives the single exciton condensation phase. If we think of the layer index as a $SU(2)$ pseudspin,  the interlayer pairing is associated with a $\vec{d}$ vector in the $z$ direction. We can write the charge $2e$ pairing as $\mathbf{\Delta_{2e}}=b_c \vec{d}$ with $b_c=e^{i\theta_c}$ as a quantum rotor carrying the charge degree of freedom. One can get a charge $4e$ pairing by keeping $\langle b_c \rangle \neq 0$ while disordering the spin part: $\langle \vec{d} \rangle =0$. This can be described as an Ising transition by prolifering $\vec{d} \rightarrow -\vec{d}$ Ising domain walls.

\section{Comments on Experiment}
We summarize our theoretical results in Fig.~\ref{fig:phase_diagram} and discuss possible experimental signatures. The primary experimental signatures traditionally used to identify the single exciton condensate in quantum wells\cite{eisenstein2014exciton} occur also in the paired exciton condensate. The paired exciton condensate is roughly equivalent to the spontaneous generation of correlated tunneling in pairs of electrons from one layer to the other, so it also shows a peak in the interlayer tunneling conductance $dI/dV$ at zero bias. Similarly for Hall resistance measurements in a counterflow geometry, with currents in opposite directions in the two layers, the current is carried by the condensed exciton pairs, which again are neutral objects, and thus the counterflow Hall resistance vanishes for both single and paired exciton condensates. 
Therefore the observation of these signatures cannot be used to distinguish the two phases. 

However scanning the phase diagram should allow for a determination of the paired exciton phase. 
When increasing the distance $d$, we expect a transition from EC to PEC at $d_{c1}$. Recall that numerical simulations for unscreened projected Coulomb interactions suggest $d_{c1}\approx 1.1 \ell_B$\cite{zhu2017numerical}.  At $d\rightarrow \infty$, the two CFLs are unstable to an inter-layer pairing, which can correspond to the same EC phase\cite{sodemann2017composite}. As a result, in this scenario there must be a second critical point $d_{c2}$ separating the intermediate PEC and the EC at large $d$, though its precise value is not clear. At large $d$, both the phase stiffness for the exciton condensation and the charge gap should be small. Therefore under disorder and at finite temperature, the EC phase may not show up in the experiments at large $d$, though signatures may become clearer with better samples.   More experimental data  in the intermediate region is needed to determine the phase boundary.

Finite temperature behavior associated with the BKT transition can point to the presence of the new PEC phase. 
Close to $d_{c1}$, there should be a finite temperature Ising transition separating EC and PEC at $T_{Ising}$ below the BKT transition at $T_{BKT}$. This thermal Ising transition onsets at $d_{T1}<d_{c1}$. This point $d_{T1}$ may be accessible in future experiments on graphene double layers. In addition, close to the tri-critical point at $d_{T1}$, the  exponent of the $I-V$ curve should show behavior markedly different from the single BKT transition, since for temperature below  $T_{BKT}$ there is a rapid crossover to critical behavior associated with the lower Ising criticality and thus an expected change in the finite temperature $I-V$ behavior away from the BKT universality.

Finally let us discuss consequences of considering the superfluid density or phase stiffness. Both the EC and PEC phases are described by a boson field: $\rho^s_{EC} (\upsilon_{EC}\partial \varphi_{EC}-A_s)^2$ for the EC and $\rho^s_{PEC} (\upsilon_{PEC} \partial \varphi_{PEC}-2A_s)^2$ for the PEC. Therefore, $\rho^s_{EC}=\rho_s$ in the EC phase and $\rho^s_{PEC}=\frac{1}{4}\rho_s$ in the PEC phase. Here $\rho_s$ is the physical phase stiffness corresponding to $\rho_s A_s^2$. At zero $T$, the transitions at both $d_{c1}$ and $d_{c2}$ should be continuous. Therefore we expect that the physical phase stiffness $\rho_s$  is smooth across the critical point. So the phase stiffness of the bosonic field drops down after entering the PEC phase from the EC phase. As a result, $T_{BKT}$ should also drop down because $T_{BKT}$ is decided mainly by the phase stiffness $\rho^s_{EC}$ or $\rho^s_{PEC}$ of the bosonic field, instead of the $\rho_s$.  Therefore a quick decrease of $T_{BKT}$ close to $d_{c1}$ is expected.

 \section{Conclusion}
 In conclusion we propose a new paired exciton phase (PEC) at intermediate interlayer distance for the bilayer quantum Hall system with $\nu_T=\frac{1}{2}+\frac{1}{2}$. We show that the elementary excitation is a half meron with $1/4$ charge and self statistics $\theta_s=\frac{\pi}{16}$. We argue that such a phase can be understood in a dual language by charge $4e$ pairing of  the two composite Fermi liquids near the $d=\infty$ limit.  We suggest experiments to look for a thermal Ising transition between EC and PEC at $T_{Ising}$ below the BKT transition close to the first critical point $d_{c1}$.

The proposed PEC phase also offers a good platform to study topological charge $4e$ superconductivity. A real charge $4e$ superconductor of electrons is usually thought to be difficult to realize in solid state systems. Our understanding of topological charge $4e$ superconductivity is quite incomplete due to the lack of the mean field theory, but the intermediate region of quantum Hall bilayers provides a good opportunity to study  topological charge $4e$  pairing of composite fermions and its transition to a charge $2e$ pairing at zero temperature.

The phase diagram we propose is also interesting from the purely theoretical view. Theoretically we expect that the transitions between single exciton condensation phase and the paired exciton condensation at both $d_{c1}$ and $d_{c2}$ can be continuous. The first transition at $d_{c1}$ is described by a superfluid to paired superfluid transition for composite bosons, while the second transition at $d_{c2}$ is described by a charge $2e$ pairing to charge $4e$ pairing transition for composite fermions. The existence of the two different descriptions for the same phase transition is a quite non-trivial demonstration for the duality between compsite boson and composite fermion.

\section{Acknowledgement}
We thank T. Senthil and Max Metlitski for useful discussions. We thank Zhu Zheng for very helpful explanations of his numerical result. YHZ is supported by NSF grant DMR-1608505 to Senthil Todadri.  IK acknowledges support from an MIT Pappalardo fellowship in physics and a CU Boulder Center for Theory of Quantum Matter Fellowship.

\bibliographystyle{apsrev4-1}
\bibliography{Pexciton}

\onecolumngrid
\appendix

\section{Hilbert Space of Single Exciton\label{appendix:exciton_eigen_states}}
One key point to write down a theory for exciton is the observation that single exciton state is labeled by a two dimensional momentum $\mathbf k$. Single exciton state is just the eigenstate of the Schrodinger equation with two oposite charges in the same magnetic field:
\begin{equation}
 \left( \frac{1}{2m} (-i\hbar\partial_{\mathbf{x_1}}-e\mathbf A(\mathbf{x_1}))^2+\frac{1}{2m} (-i\hbar\partial_{\mathbf{x_2}}+e\mathbf A(\mathbf{x_2}))^2\right)\Psi(\mathbf{x_1},\mathbf{x_2})=E \Psi(\mathbf{x_1},\mathbf{x_2})
 \label{eq:opposite_charge_Landau}
\end{equation}
where $\mathbf{x_1}$ is the coordinate of the electron for layer $1$ and $\mathbf{x_2}$ is the coordinate of the hole for layer $2$.

Because two particles are not interacting, one can easily solve the standard Landau level problem for the electron and the hole separately. The eigenstate of exciton is then labeled by $mn$ with wavefunction $\Psi_{mn}(\mathbf{x_1},\mathbf{x_2})=\varphi_{m}(\mathbf{x_1})\bar{\varphi}_n(\mathbf{x_2})$. Here $\varphi_m(\mathbf{x_1})$ is the wavefunction for the $m$ th  Landau orbital and $\bar{\varphi}_n(\mathbf{x_2})$ is the wavefunction of the $n$th Landau orbital for an opposite charge.

Two Landau orbital index $mn$ makes the Hilbart space of the single exciton complicated. However, there is an equivalent way to solve the Eigenstates for Eq.~\ref{eq:opposite_charge_Landau}. One can use the center-of-mass coordinate: $\mathbf{R}=\frac{\mathbf{x_1}+\mathbf{x_2}}{2}$ and $\mathbf{r}=\mathbf{x_1}-\mathbf{x_2}$. Here $\mathbf{R}$ is the coordinate of the exciton and $\mathbf{r}$ is the dipole moment. Because exciton is neutral, the eigenstate can be labeled by well-defined center of mass momentum.  By solving the problem in the center-of-mass coordinate system, we get the eigenfunction for center-of-mass momentum $\mathbf k$, which is the $\tau_{\mathbf{k}}(z,\bar w)$ in Eq.~\ref{eq:exciton_eigen_states}. As expected the dipole moment is locked to $\mathbf{k}$. Because we just change the coordinate system, Eq.~\ref{eq:complete_condition} naturally follows. Therefore we can use $\mathbf{k}$ to label the single exciton state, which is used to write down an effective theory for excitons in this paper. This wavefunction $\tau_{\mathbf k}(z,\bar w)$ was actually proposed before By Pasquier-Haldane\cite{pasquier516dipole} and N.Read\cite{read1998lowest}. In these previous papers the opposite charge is an auxiliary  particle which is not physical and essentially  the physical Hilbert space is enlarged. In contrast, in this paper both particles are physical and we are trying to restrict to the Hilbert space of exciton.

\section{Self Consistent Bogoliubov Mean Field Theory\label{appendix:self-consistent-equation}}
We derive the self consistent equations for both superfluid and paired superfluid phase for the bosonic excitons following the standard Bogoliubov theory.

\subsection{Single Exciton Condensation}
We do the standard Bogoliubov mean field theory to describe the phase with the single exciton condensed.
\begin{equation}
  H_M=H_C+H_0+H_V
\end{equation}
where $H_C$ is a constant energy term for condensation:
\begin{equation}
  H_C=V_{+}(0)\left( \frac{\Phi^4+2\Phi^2 \sum_{q\neq 0}(b_q^\dagger b_q)}{V^2}\right) \sim V_{+}(0) n^2
\end{equation}
where $n=\frac{N}{V}$ and $\Phi=\langle b \rangle$.  From mean field decoupling of interaction term in Eq.~\ref{eq:four-boson} we get:
\begin{equation}
  V_{+}(\mathbf q)=\frac{e^{-\frac{|q|^2}{2}}}{|q|}(1-e^{-|q|d})
\end{equation}
which implies that
\begin{equation}
  V_{+}(0)=d
\end{equation}

$H_0$ is the $b^\dagger b$ term:
\begin{equation}
  H_0=\sum_{q\neq 0}(\xi(\mathbf q)+2 n_0 V_{+}(\mathbf q)) b^\dagger_{\mathbf q} b_{\mathbf q}
\end{equation}
where $n_0=\Phi^2$ and $\xi(\mathbf q)=\frac{|\mathbf q|^2}{2m}$.

We also have a $b b$ term
\begin{equation}
  H_V=\sum_{\mathbf q \neq 0} n_0 V_{+}(\mathbf q)(b^\dagger_{\mathbf q} b^\dagger_{-\mathbf q}+b_{\mathbf q} b_{-\mathbf q})
\end{equation}

We can diagonalize the Hamiltonian using
\begin{equation}
  \alpha_{\mathbf q}=\mu_{\mathbf q}b_{\mathbf q}+\upsilon_{\mathbf q}b^\dagger_{-\mathbf q}
\end{equation}

with
\begin{align}
\mu_{\mathbf q}^2&=\frac{1}{2}\left(\frac{\xi_{\mathbf q}+2 n_0V_{+}(\mathbf q)}{E_{\mathbf q}}+1\right)\notag\\
\upsilon_{\mathbf q}^2&=\frac{1}{2}\left(\frac{\xi_{\mathbf q}+2 n_0V_{+}(\mathbf q)}{E_{\mathbf q}}-1\right)
\end{align}
where $\xi_{\mathbf q}=\frac{|\mathbf q|^2}{2m}$ and $E_{\mathbf q}=\sqrt{(\epsilon_{\mathbf q}+2 n_0V_{+}(\mathbf q))^2-(2 n_0V_{+}(\mathbf q))^2}$.

From the condition $n_0=n-\sum_{\mathbf q \neq 0}\langle b^\dagger_{\mathbf q}b_{\mathbf q}\rangle$ we get the self consistent equation in Eq.~\ref{eq:self_mean}.

\subsection{Paired Exciton Condensation}
For paired superfluid phase of the bosonic excitons, we need to add a non-zero chemical potential $\mu$.  We decouple the four boson interaction in Eq.~\ref{eq:four-boson} to

\begin{equation}
  H_{M}=\sum_{\mathbf k} (\xi(\mathbf k)+A(\mathbf k)-\mu)b^\dagger_{\mathbf k} b_{\mathbf k}+\left(B(\mathbf k) b_{\mathbf k} b_{-\mathbf k} +h.c.\right)
\end{equation}

with self consistent equations:
\begin{equation}
  A(\mathbf k)=\frac{1}{V}\sum_{\mathbf q} \frac{1}{q}e^{-\frac{q^2}{2}}\left(1-e^{-qd}\cos(\mathbf k \wedge \mathbf q)\right)\langle b^\dagger_{\mathbf{k+q}}b_{\mathbf{k+q}} \rangle
\end{equation}

\begin{equation}
  B(\mathbf k)=\frac{1}{V}\sum_{\mathbf q} \frac{1}{q}e^{-\frac{q^2}{2}}\left(\cos(\mathbf k \wedge \mathbf q)-e^{-qd}\right)\langle b^\dagger_{\mathbf{k+q}}b^\dagger_{-(\mathbf{k+q)}} \rangle
\end{equation}
and
\begin{equation}
  n=\frac{1}{V}\sum_{\mathbf k}\langle b^\dagger_{\mathbf k}b_{\mathbf k}\rangle
\end{equation}

\end{document}